# Single-shot diffraction-limited imaging through scattering layers via bispectrum analysis

**Tengfei Wu,[1,2] Ori Katz,[1,3,*] Xiaopeng Shao,[2] Sylvain Gigan [1]**

[1] *Laboratoire Kastler Brossel, ENS-PSL Research University, Paris 75005, France*
[2] *School of Physics and Optoelectronic Engineering, Xidian University, Shaanxi 710071, China*
[3] *Department of Applied Physics, The Hebrew University of Jerusalem, Jerusalem 9190401, Israel*

**Corresponding author: orik@mail.huji.ac.il*

**Recently introduced speckle-correlations based techniques enable noninvasive imaging of objects hidden behind scattering layers. In these techniques the hidden object Fourier amplitude is retrieved from the scattered light autocorrelation, and the lost Fourier phase is recovered via iterative phase-retrieval algorithms, which suffer from convergence to wrong local-minima solutions and cannot solve ambiguities in object-orientation. Here, inspired by notions used in astronomy, we experimentally demonstrate that in addition to Fourier amplitude, the object phase information is naturally and inherently encoded in scattered light bispectrum (the Fourier transform of triple-correlation), and can also be extracted from a single high-resolution speckle pattern, based on which we present a single-shot imaging scheme to deterministically and unambiguously retrieve diffraction-limited images of objects hidden behind scattering layers.**

The inherent inhomogeneity of complex samples encountered in many imaging scenarios induces light scattering, which diffuses the light from any object buried inside or hidden behind such samples into a complex speckle pattern [1], and makes direct imaging of such objects impossible. In recent years, several approaches have been put forward to overcome this seemingly intractable problem. Wavefront shaping has been demonstrated as a powerful technique for focusing and imaging through scattering media [2-11]. However, these methods require a detector or an optical/acoustical probe in the plane of interest. These techniques are time-consuming due to the required long sequence of measurements steps, one for each imaging pixel, and are thus difficult to use with dynamic samples. A recent breakthrough approach by Bertolotti et al. [12], which exploits the inherent angular correlations of scattered light speckle patterns, known as the *'memory effect'* [13, 14] has realized imaging through scattering layers without any guide star or complex wavefront shaping process. However, this technique requires a long angular-scanning acquisition sequence, which limits its use to relatively static samples. Based on the same speckle-correlations and inspired by the astronomical technique of stellar speckle interferometry, Katz et al. [15] have demonstrated that objects hidden behind scattering layers can be retrieved from the autocorrelation of a single high-resolution scattered light image, captured by a standard camera, via iterative phase-retrieval algorithms [16, 17]. The single-shot technique benefits from a short acquisition time, which makes it possible to realize real-time imaging. However, this technique suffers from some shortcomings of iterative phase-retrieval algorithm: i). it requires large numbers of iterations and independent runs with different random initial guesses to avoid stagnating at local optimal solution; ii). The convergence of phase-retrieval algorithm depends on the prior information (e.g. the accuracy of the estimated Fourier amplitude of object) or some assumptions (e.g. the size of support area [16]).

In this Letter we experimentally demonstrate a single-shot noninvasive imaging scheme for realizing diffraction-limited observation of hidden objects behind scattering layers, without the use of iterative phase-retrieval algorithms. Inspired by techniques used in astronomy and based on the notion of closure-phase pioneered in radio-astronomy, we extract the object's Fourier phase deterministically and unambiguously via bispectrum (triple-correlation) analysis of a single scattered light pattern. Just as in Katz's et al. [15] work only a single camera image is required. In addition to being deterministic and straightforward to implement, the technique benefits from the reduced sensitivity of bispectrum to additive Gaussian noise [18], which is an important practical advantage.

The schematic of the experimental setup and a numerical simulation of our imaging scheme are presented in Fig. 1. In Fig. 1(a), an object that is placed at a distance $u$ behind a highly scattering medium, is illuminated by a spatially incoherent light. A high-resolution camera is placed at a distance $v$ from the other side of the scattering medium to capture the transmission light. If the object lies within the range determined by the optical memory effect, points on the object will produce nearly identical, but shifted random speckle patterns. For spatially incoherent illumination, the camera image is a simple superposition of these random speckle patterns, which allows the system in Fig. 1(a) to be treated as an incoherent imaging system with a space–invariant point spread function (PSF), i.e. the identical and shifted random speckle pattern. With the magnification of $M=v/u$, the camera image $I$ [Fig. 1 (b)] is a convolution of the object intensity pattern and the PSF:

$$I(v\theta) = O(u\theta) * S(\theta) \tag{1}$$

where $*$ denote the convolution operator. $S(\theta)$ is the PSF, and $\theta$ is the viewing angle. The Fourier amplitude of object [Fig. 1 (c)] can be extracted from the autocorrelation of the single high-resolution camera image. We refer the reader to [15] for more information on this step. The Fourier phase of object [Fig. 1 (d)] is recovered separately and independently from the bispectrum analysis of the large speckle pattern (see details below and in Fig. 3). The final imaging result [Fig. 1(e)] is achieved by a simple inverse Fourier transform of the combination of the estimated Fourier amplitude and Fourier phase.

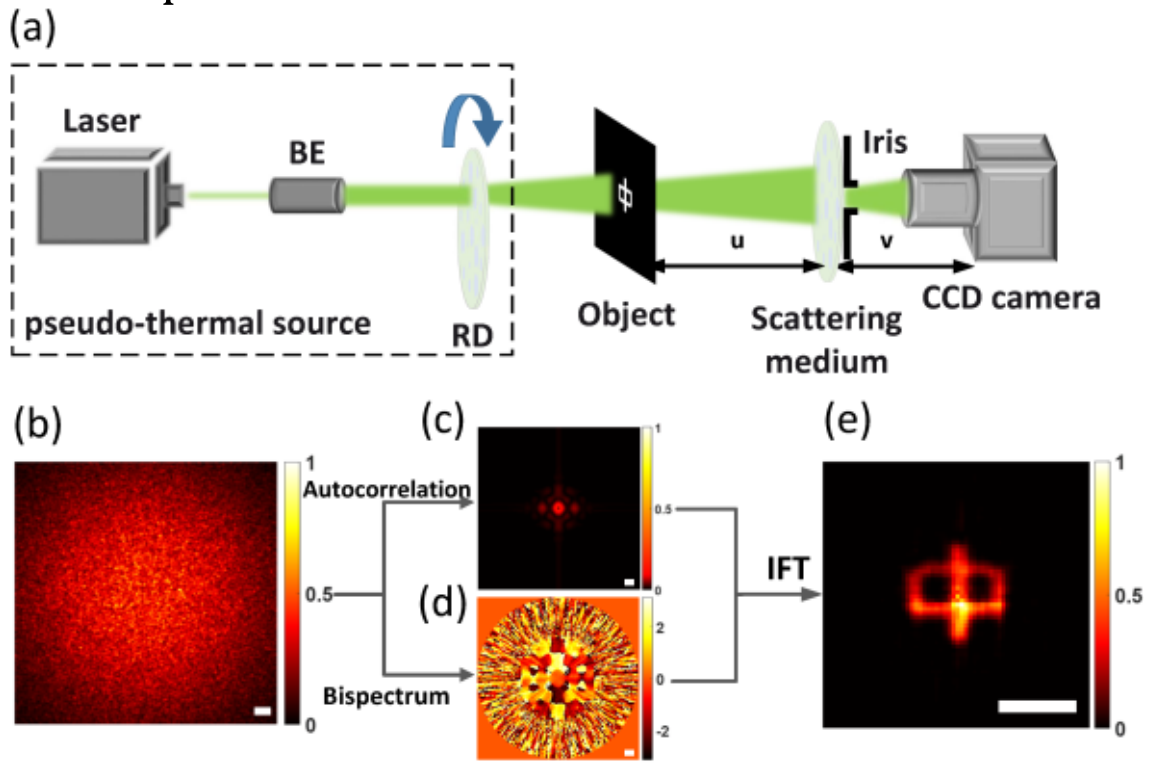

Fig. 1 Single-shot and noninvasive imaging through scattering layers with bispectrum analysis: experimental setup and numerical simulation. (a) experimental setup: BE: beam expander; RD: rotating diffuser. (b) high-resolution camera image. Estimated (c) Fourier amplitude and (d) Fourier phase. (e) final imaging result displayed in intensity scale. Scale bar: 150 pixels in (b) and 20 pixels in (c), (d) and (e).

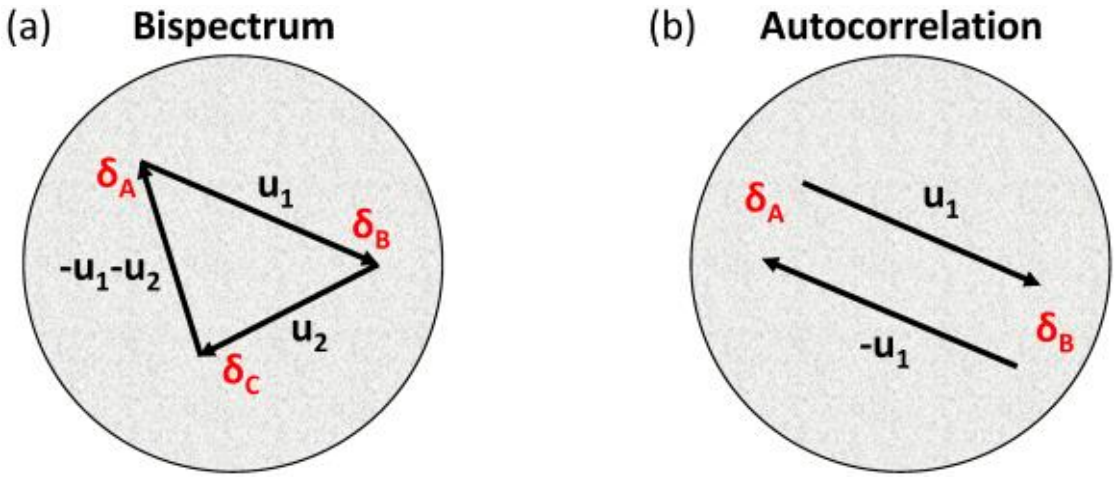

Fig. 2 Schematic of closure-phase theory for (a) bispectrum and (b) autocorrelation.

Bispectrum is a generalization of closure-phase theory employed in radio astronomy [19]. Considering the telescope aperture in Fig. 2(a), in which $\delta_A$, $\delta_B$ and $\delta_C$ are three constant, but random phases, caused by atmosphere turbulence, in three uncorrelated sub-apertures. The three spatial frequencies, $u_1$, $u_2$ and $-u_1$-$u_2$, determined by the three sub-apertures form a closed triangle. Assuming an object is imaged through such a three-sub-aperture telescope, one would obtain three superimposed fringe patterns. The bispectrum of the fringe patterns $R$ is defined as [20]:

$$B_R(u_1, u_2) = \tilde{R}(u_1) \cdot \tilde{R}(u_2) \cdot \tilde{R}(-u_1 - u_2) \tag{2}$$

where $\tilde{R}(\cdot)$ denotes the Fourier transform of the fringe patterns. For the phase term of Eq. (2):

$$\begin{aligned} \Phi\left\{B_R(u_1, u_2)\right\} &= \Phi\left\{\tilde{R}(u_1) \cdot \tilde{R}(u_2) \cdot \tilde{R}(-u_1 - u_2)\right\} \\ &= \Phi\left\{\tilde{O}(u_1)\right\} + \delta_A - \delta_B \\ &\quad + \Phi\left\{\tilde{O}(u_2)\right\} + \delta_B - \delta_C \\ &\quad + \Phi\left\{\tilde{O}(-u_1 - u_2)\right\} + \delta_C - \delta_A \\ &= \Phi\left\{\tilde{O}(u_1)\right\} + \Phi\left\{\tilde{O}(u_2)\right\} + \Phi\left\{\tilde{O}(-u_1 - u_2)\right\} \end{aligned} \tag{3}$$

where $\Phi\{\cdot\}$ is the phase information and $\tilde{O}(\cdot)$ denotes the Fourier transform of object. Eq. (3) indicates that the bispectrum preserves the phase information of object. With similar derivation, we can find the autocorrelation technique, which corresponds to the power spectrum in Fourier space [Fig. 2(b)], retains only the amplitude information.

In order to extract the phase information from the bispectrum, the statistical noise of bispectrum should be suppressed adequately [21]. Benefiting from the high pixel count of modern digital camera and from the ergodicity of the scattering process, we can extract the Fourier phase of object from the bispectrum analysis of a single high-resolution speckle pattern, by exploiting the concept of replacing temporal average with spatial average. We divide the single camera image into multiple sub-images $I'$, and filter each one with a Gaussian window function in real space [Fig. 3(a)]:

$$G = \exp\left[-\left(x^2 + y^2\right)/w^2\right] \tag{4}$$

where x and y are the coordinates in real space, and $w$ denotes the window size. In general, $w$ should be larger than twice of the object size in one dimension, which can be roughly and independently estimated by its autocorrelation. An overlap should be conserved between adjacent sub-images to make sure we maximally exploit the full information contained in the single camera image.

To avoid the huge computational complexity caused by the 4D bispectrum of each 2D image, we utilize the projection method in [22] and project the speckle patterns into a series of 1D projections with the angle uniformly distributed from 0 to π. The bispectrum of the $p$-th 1D projection is calculated by Eq. (2). Owing to convolution theorem, the averaged bispectrum of 1-D projection of each sub-image is given by:

$$\left\langle B_{I'_p} \right\rangle_n = B_{O_p} \cdot \left\langle B_{S'_p} \right\rangle_n \tag{5}$$

where $\langle \cdot \rangle$ is the average operator and $n$ is the number of sub-image. $I'_p$, $O_p$ and $S'_p$ are the $p$-th projection of the sub-image, the object and the spatially ergodic "sub-PSF". As the temporally averaged bispectrum of PSF, its spatial average $\left\langle B_{S'_p} \right\rangle_n$ in our case has real values, which deduces the following relationship:

$$\Phi\left\{\left\langle B_{I'_p} \right\rangle_n\right\} \approx \Phi\left\{B_{O_p}\right\} \tag{6}$$

Eq. (6) shows that the bispectrum analysis of a single high-resolution speckle pattern enables us to obtain the bispectrum phase of object [Fig. 3(b)]. Once the bispectrum phase of object is obtained, a recursive process as expressed in Eq. (7) can be used to extract the Fourier phase of object:

$$\Phi\left\{\tilde{O}_p\left(-u_1-u_2\right)\right\}=-\Phi\left\{\tilde{O}_p\left(u_1\right)\right\}-\Phi\left\{\tilde{O}_p\left(u_2\right)\right\} +\Phi\left\{B_{O_p}\left(u_1,u_2\right)\right\} \quad (7)$$

We assume the initial condition as $\Phi\left\{\tilde{O}_p\left(0\right)\right\}=\Phi\left\{\tilde{O}_p\left(1\right)\right\}=0$, and the recursive algorithm then proceeds to higher frequencies. We should also note that the bispectrum allows reconstructing the object-phase, up to a gradient, therefore the absolute position of the object is undetermined. Here, the above assumption amounts to fix the position of the object, but has no influence on its structure [20]. As an example, the recovered 1D Fourier phase from the bispectrum phase, corresponding to the projection $\theta = \pi/3$, is presented in the dashed box in Fig. 3(c). To examine the validity of the recovered 1D Fourier phase, we compare the reconstructed 1D projection (solid line) and the true projection of object (dashed line) in Fig. 3(c). Since the bispectrum preserves the true and completed object phase information, which mainly determines the structure of an image, we can also obtain the true object-orientation by extracting the phase information from bispectrum analysis.

The central slice theorem states that the Fourier transform of a projection of object at a given angle is same as the parallel central section through the Fourier transform of that object [23]. In our case, we arrange each recovered 1D Fourier phase to the position determined by the projection angle in 2D Fourier space [Fig. 3(c)]. Triangulation-based cubic interpolation is implemented to solve the problem of coordinate transformation from Polar coordinate system to Cartesian coordinate system. The final 2D Fourier phase of object is achieved after arranging all the recovered 1D Fourier phases to the corresponding positions [Fig. 3(d)].

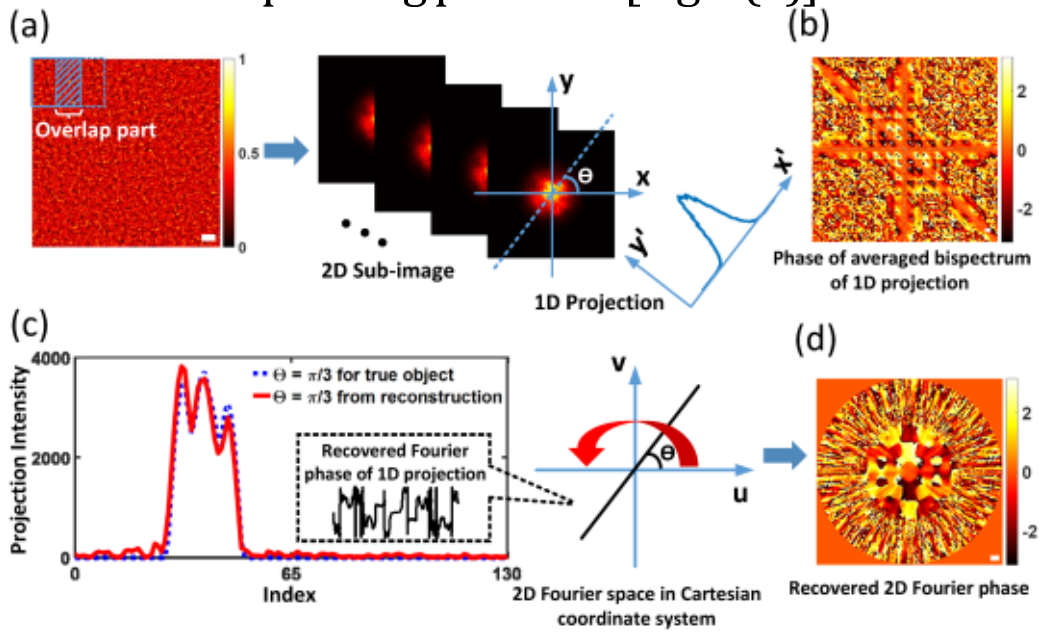

Fig. 3 Detailed process of extracting the phase information from single camera image. (a) pre-processed high-resolution camera image and the divided multiple 2D sub-images. (b) phase of the averaged bispectrum of 1D projection (θ=$\pi$/3) of sub-images. (c) comparison of the 1D recovered projection of object (red solid line) from single camera image with the true projection of object (blue dashed line). The recovered 1D Fourier phase (in the black dashed box) is arranged to the 2D Fourier space. (d) 2D Fourier phase of object from bispectrum analysis. Scale bar: 150 pixels in (a) and 20 pixels in (b) and (d).

In the experimental demonstrations of this concept, the light source is a spatially incoherent and narrowband pseudo-thermal source, composed by a 532nm single frequency CW laser (Cobolt Samba™-100), whose beam diameter is expanded by a 10× beam expander, and a rapidly rotating diffuser. The incoherent light illuminates the object and the transmission light reaches the scattering medium (Edmund, Ground Glass Diffuser), which is placed ~60 cm in front of the object. An iris with a diameter of ~0.3cm is positioned against the other side of the scattering medium. After passing through the scattering medium, the transmission light is captured by a high-resolution camera (Andor, ZYLA-5.5-USB3, 2160×2560), which is placed ~12cm in front of the scattering medium. The objects in this experiment are digits 1, 4 and 5 from the United States Air Force resolution target (Edmund, 1951 USAF Negative Target, 2"×2", Group "1").

The raw camera images [Fig. 4(a)] are spatially normalized for the slowly varying envelope of the scattered light pattern halo. The normalized images are smoothed by a Gaussian kernel [15]. The Fourier amplitudes in Fig. 4(b) are recovered from the autocorrelation of each pre-processed camera image. We divide the pre-processed camera image into multiple sub-images, each of which is filtered by the Gaussian window function of Eq. (4) with a width of 50 pixels. The overlap part between adjacent sub-images is around 120 pixels. We project each sub-image into 180 projections via a Radon transform and calculate the averaged 2D bispectrum of each 1D projection. In our case, we use the Higher Order Spectral Analysis Toolbox [24], which is an external Matlab package, to calculate the bispectrum. The Fourier phase of each 1D projection is extracted from the 2D bispectrum with the recursive algorithm. The Fourier phase of objects [Fig. 4(c)] are achieved by arranging all the 1D Fourier phase of 180 projections into the corresponding positions of 2D Fourier space. After implementing the inverse Fourier transform, the final imaging results are achieved [Fig. 4(d)]. Although 180 projections were used in our experiments, it is important to note that for the relatively simple objects used, a few tens of projections are sufficient for reasonable reconstruction, which is beneficial for reducing the computational complexity. As an example, we use 18 projections [Fig. 5(a)] and 9 projections [Fig. 5(b)] to reconstruct the speckle pattern of digit 4. Both of them work expect for the fact that more projections (i.e. more samplings in the Fourier domain), smoother the reconstruction is. However, insufficient projection, as shown in Fig. 5(c), would lose the necessary phase information.

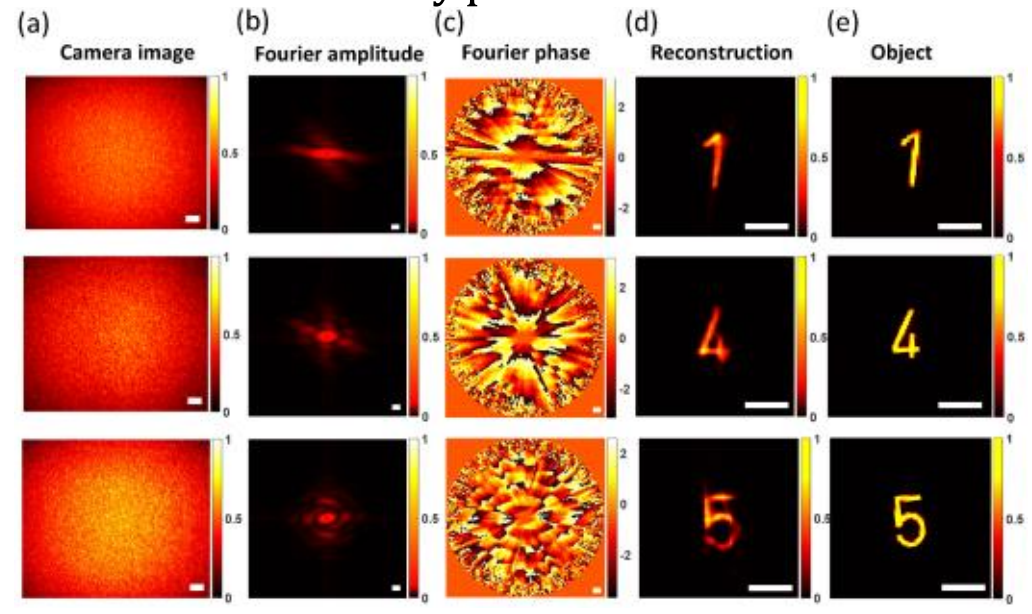

Fig. 4 Experimental results of imaging through an opaque ground diffuser. Column (a): raw camera images. (b) and (c) are the estimated Fourier amplitude and Fourier phase. (d): images of the hidden objects (display in intensity). (e) is the objects. Scale bar: 150 pixels in column (a) and 20 pixels in the others.

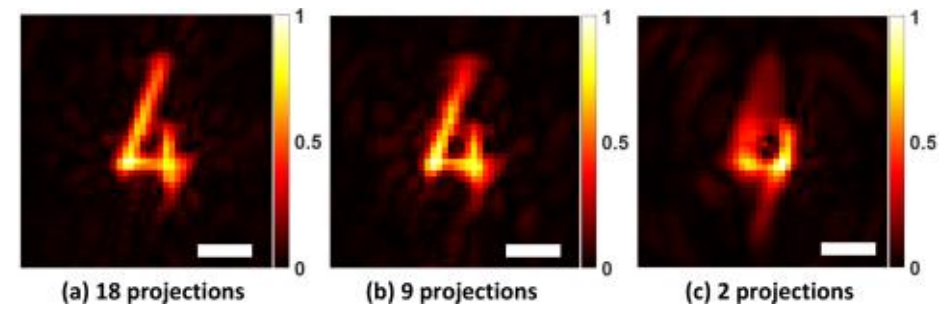

Fig. 5 Reconstructions with (a) 18 projections, (b) 9 projections and (c) 2 projections (display in amplitude). Scale bar: 10 pixels.

As presented, the whole reconstruction process in our imaging scheme is fully deterministic and non-iterative. The other significant superiority of our approach is the reduced sensitivity to experimental noise. In general, the phase estimation from phase-retrieval depends on the accuracy of the estimation of Fourier amplitude of object, which can be extracted well from the speckle pattern in "clean" experimental environments. In our tests with the basic phase-retrieval algorithms, i.e. the combination of Hybrid Input-Output (HIO) and Error Reduction (ER) algorithms [12, 15, 16], the reconstruction always fails in noisy cases, because of the noisy Fourier amplitude estimation. Interestingly, our approach suffers much less in noisy environments, since phase extraction from bispectrum analysis is relatively insensitive to additive Gaussian noise [18]. Owing to the fact that phase information preserves more image features than amplitude, acceptable imaging results can still be achieved in noisy environments with our method.

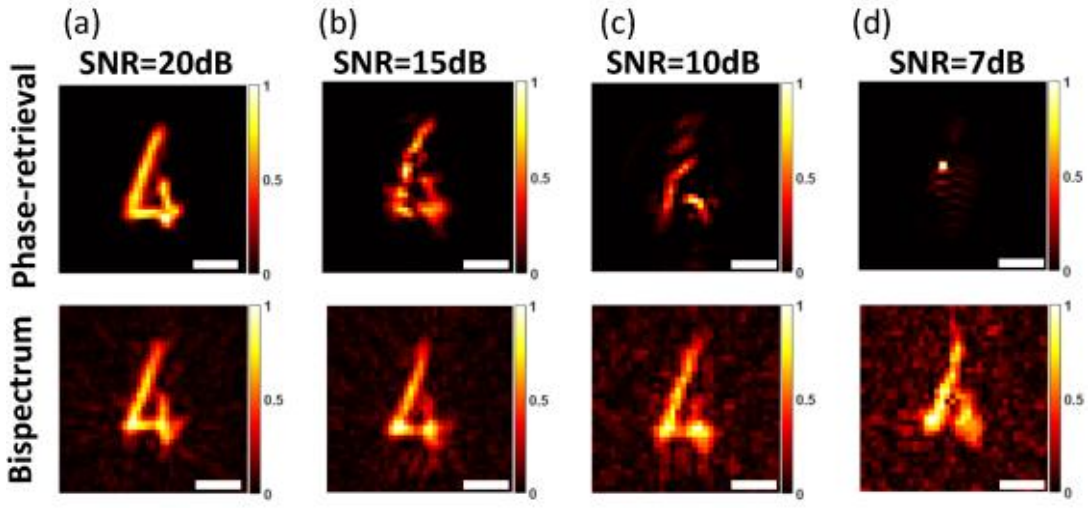

Fig. 6 Performance of basic phase-retrieval and bispectrum in noisy environments (simulated noise). Top row shows the results of basic phase-retrieval based method. Bottom row presents the results of bispectrum method. All the results are shown in amplitude. Scale bar: 10 pixels.

To demonstrate this, we add four different levels of white Gaussian noise, which is a common-used model of experimental noise (e.g. read noise of camera), to the speckle pattern for digit 4. The signal-to-noise ratio (SNR) for each artifactitious speckle pattern is 20dB, 15dB, 10dB and 7dB. For phase-retrieval reconstruction, we run 1530 iterations of HIO with a decreasing $\beta$ factor from 2 to 0, with a step of 0.04. The result of HIO is then used as an initial guess to additional 30 iterations of ER to achieve the final result. For each noisy case, phase-retrieval reconstruction is performed with 200 different random initial guesses and the convergence is monitored by the mean square error (MSE) metric between the Fourier amplitude of the reconstructions and the estimated Fourier amplitude of object from the original "clean" speckle pattern:

$$E = \frac{\sum_{i=1}^{n}\left[\left|\mathcal{F}\{R\}\right|-\left|\mathcal{F}\{O\}\right|\right]^2}{n} \quad \textbf{(8)}$$

where $\mathcal{F}\{\cdot\}$ denotes the Fourier transform. $R$ and $O$ represent the reconstruction and the object, $n$ is the number of elements. The trials with lowest MSE in each case are chosen as the final imaging results [top row in Fig. 6]. In weak noise case (SNR=20dB), although phase-retrieval can perform as well as bispectrum, it needs many independent runs to avoid local minimum, and the fluctuation denoting the imaging quality of different trials would be stronger even in this weakly noisy case. With the increase of noise, the imaging result of phase-retrieval becomes distorted (SNR=15dB), and eventually unrecognizable (SNR=10dB and SNR=7dB). For bispectrum, the main influence of noise is only to increase the background intensity, while the contrast between the signal and the background is high enough to observe the object. Actually, since the bispectrum of Gaussian processes is identically zero [18], the influence of the additive Gaussian noise can be further suppressed by increasing the number of speckle grains, which can lead to better ensemble averaging of bispectrum.

In conclusion, we experimentally demonstrated that in addition to the Fourier amplitude, the Fourier phase of object is inherently and naturally encoded in the scattered light bispectrum and we can deterministically and unambiguously extract the accurate object phase information from the bispectrum analysis of a single camera image of scattered light, based on which we present a noninvasive single-shot imaging scheme to observe the hidden objects through scattering layers. Compared to phase-retrieval based methods, our imaging scheme not only is deterministic and non-iterative, but also can solve ambiguities in object-orientation and performs better in noisy situations, which make it more attractive for many potential areas of applications, e.g. bio-imaging or optical detection. For this proof of principle, we only used here the basic recursive algorithm [20], many other advanced algorithms are expected to perform much better in extracting phase information of object from bispectrum [25, 26]. An interesting point of our method is to use the phase solution from bispectrum analysis as the initial condition of phase-retrieval, which can make the phase-retrieval process converge very fast to the optimized solution and also obtain good imaging results. To this end, we freely make available the source codes and experimental data to use by the scientific community, as shown in Codes and Data file (Ref. [27]).

**Funding.** European Research Council (278025 and 677909); China Scholarship Council (201606960026); National Natural Science Foundation of China (61575154); Azrieli Foundation.

**Acknowledgment**.
The authors would like to thank Thomas Chaigne and Jonathan Dong for the useful discussions, and Huijuan Li for the help of experiments.